\documentclass[aps,pre,twocolumn,groupedaddress,showpacs]{revtex4-1}
\usepackage{amsmath,amssymb,fancybox,txfonts,bm,color}
\usepackage[dvipdfmx]{graphicx}
\usepackage{braket}
\begin{document}
\title{
Lotka--Volterra Competition Mechanism Embedded in a Decision-Making Method
}

\newcommand{\tP}{\mbox{$\bar{P}$}}

\newcommand{\niyama}[1]{\textcolor[rgb]{1,0,0}{#1}}
\newcommand{\ave}[1]{ \left< #1 \right>}
\newcommand{\vc}{\mathbf}
\newcommand{\gvc}[1]{\mbox{\boldmath $#1$}}
\newcommand{\emf}[1]{{\gtfamily \bfseries #1}}
\newcommand{\fracd}[2]{\frac{\displaystyle #1}{\displaystyle #2}}
\newcommand{\del}[3] {\frac{\partial^{#3} #1}{\partial #2^{#3}}}
\newcommand{\dev}[3]{\frac{\text{d}^{#3} #1}{\text{d}#2^{#3}}}
\newcommand{\pdev}[3]{{\text{d}^{#3} #1}/{\text{d}#2^{#3}}}
\newcommand{\intd}[1]{\text{d} {#1}}



\author{Tomoaki Niiyama$^{1}$}
\email{niyama@se.kanazawa-u.ac.jp}
\author{Genki Furuhata$^{2}$}
\author{Atsushi Uchida$^{3}$}
\author{Makoto Naruse$^{4}$}
\author{Satoshi Sunada$^{1, 5}$}

\affiliation{
$^{1}$
Faculty of Mechanical Engineering, Institute of Science and
Engineering, Kanazawa University,
Kakuma-machi Kanazawa, Ishikawa 920-1192, Japan \\
$^{2}$
Graduate School of Natural Science and Technology, Kanazawa University,
Kakuma-machi, Kanazawa, Ishikawa 920-1192, Japan\\
$^{3}$
 Department of Information and Computer Sciences, Saitama University,
 255 Shimo-Okubo, Sakura-ku, Saitama City, Saitama, 338-8570, Japan\\
$^{4}$ Department of Information Physics and Computing, Graduate School of
 Information Science and Technology, The University of Tokyo,
7-3-1 Hongo, Bunkyo-ku, Tokyo 113-8656, Japan\\
$^{5}$
Japan Science and Technology Agency (JST), PRESTO, 4-1-8 Honcho,
Kawaguchi, Saitama 332-0012, Japan
}

\begin{abstract}
Decision making is a fundamental
capability of living organisms, and has recently been gaining increasing importance in many engineering
applications. 
Here, we consider a simple decision-making principle to
identify an optimal choice in multi-armed bandit (MAB) problems, which
is fundamental in the context of reinforcement learning. 
We demonstrate that the identification mechanism of the method 
is well described by using a competitive ecosystem model, i.e., 
the competitive Lotka--Volterra (LV) model. 
Based on the ``winner-take-all'' mechanism in the competitive LV model, 
we demonstrate that non-best choices are {\it eliminated} 
and only the best choice {\it survives}; the failure of
the non-best choices exponentially decreases while repeating the choice
trials. 
Furthermore, we apply a mean-field approximation to the
proposed decision-making method and show that the method has 
an excellent scalability of $O(\log N)$ with respect to the number of choices $N$.
These results allow for a new perspective on optimal search
capabilities in competitive systems.
\end{abstract}

\maketitle

\section{Introduction}

Recent research suggests that nature is a great source 
of inspiration in providing solutions for complicated problems and
developing intelligent information processing \cite{Agarwal2014naturalComputing}. 
Inspired by biological functions, physical structures, and organizational principles found in nature, numerous mathematical and metaheuristic models have been developed. 
These include genetic algorithms, ant colony optimization, bee algorithms, and simulated annealing, which have been used for addressing various optimization problems \cite{Agarwal2014naturalComputing,Holland1989GA,Dorigo1996AntColony,Karaboga2007BeeAlgorithm,Karaboga2008BeeAlgorithm,Kirkpatrick1983SimulatedAnnealing,Cerny1985SimulatedAnnealing}.

Nature-inspired algorithms have also been applied to 
solve the {\it multi-armed bandit (MAB) problem},
which is fundamental in the context of decision making
and reinforcement learning \cite{Sutton2018ReinforcementLearning}.
A major goal of the MAB problem is 
to identify the best choice (best arm) from the available options, i.e.,
{\it best arm identification} \cite{Audibert2010BestArmMAB}.
For this purpose, various methods have been developed
\cite{Sutton2018ReinforcementLearning,Audibert2010BestArmMAB,Kim2010TOW,Kim2015TOW}
in the face of a common difficulty; 
sufficient exploratory actions may allow us to determine the best choice
with a high confidence level, but it may be accompanied by a cost.
Among the methods, the one mentioned in the Refs. \cite{Kim2010TOW,Kim2015TOW} 
is based on an inspiration from the spatiotemporal dynamics of micro-organisms, 
such as amoebas, for identifying the best choice 
in an MAB problem.
The dynamic stretching and contracting of amoebas when seeking food while maintaining their
volume constant generates a frustrating
non-local correlation as a whole, leading to an efficient and adaptive 
ability of identifying the best choice.
At present, the amoeba-inspired decision-making method has been 
implemented in various physical systems 
\cite{Kim2010TOW,Kim2015TOW,Naruse2015SinglePhotonDM,
Kim2016harnessDM,
Naruse2017Ultrafast-photo,Homma2019DecisionMaking};
however, the theoretical guarantee of the best choice identification has not yet been provided. 

 Meanwhile, a frustration similar to that      in the amoeba dynamics, 
i.e., fluctuating dynamics under a conservative constraint, 
can generally be seen in a variety of competitive dynamical systems,
in which each component (or state) competes for common finite resources.
For instance, each species in an ecosystem attempts to grow its population while competing for limited resources.
Such interspecific competition has been modeled by simple ordinary differential
equations, known as {\it the competitive Lotka--Volterra (LV) equations}
\cite{Baigent2016Lotka-Volterra,Zeeman1995extinction}.
The LV equations describe the dynamics of competitive systems,
such as multi-mode lasers
\cite{Sargent1974LaserPhysics,Sargent1993MultiModeLaser},
as well as ecological communities \cite{Baigent2016Lotka-Volterra}.
Moreover, the LV model is closely related to the Moran process in the
context of population genetics \cite{Noble2011LV-dynamics}
and evolutionary game theory \cite{Hofbauer2003Evolutionary},
suggesting the applicability of the competitive mechanism 
to explore an optimal solution adapted to a given environment.

In this study, we propose a decision-making principle in which 
the LV competitive mechanism is embedded. 
The model is a natural and simple extension of the amoeba-inspired
decision-making method \cite{Kim2010TOW,Kim2015TOW} and enables the
identification of the best choice in the MAB problems, based on the
competitive growth under a conservation law. 
We theoretically ensure the validity of the best choice identification, 
based on the LV competition model.

\section{Model}

\begin{figure*}[tbp]
\centering
\includegraphics[width=12cm]{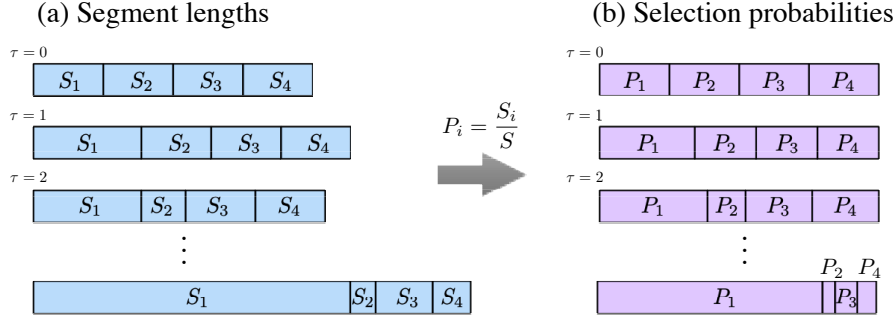}
\caption{\label{fig1} 
(Color online) Schematics of decision-making method.
(a) Lengths of the object and each segment $S_i$ 
change according to Eq.~(\ref{eq:update-S_i}).
(b) Selection probabilities $P_i$ change according to 
the change in the segment lengths.
}
\end{figure*}

First, let us consider an MAB with $N$ arms providing
unknown stochastic rewards $x_1, \cdots, x_N$, 
where $x_i$ ($i=1, \cdots, N$) is assumed to be an independent realization of a random variable 
with mean $\mu_i$.
The mean of the reward from the $i$-th arm 
is expressed as $\mu_i = \int \phi_i(x_i)x_i \intd{x_i}$,
where $\phi_i$ is the probability density of $x_i$.
Without loss of generality, we assume that $\mu_i$ is positive. 
A goal of the MAB problem is to identify the best arm with
the largest mean reward, $\mu_{i^*} = \max_i \{ \mu_i \}$, 
through multiple plays,
where $i^*$ is the best arm.
Hence, the MAB problem can be considered, for example, as a decision-making problem 
for a gambler who plays slot machines (or a slot machine with multiple arms).

Our decision-making method for identifying the best arm
is based on the dynamic behavior
of an object with a total length of $S$, which consists of $N$ segments,
as schematically shown in Fig.~\ref{fig1}(a). 
Let $S_i(\tau)$ be the length of the $i$-th segment at the $\tau$-th play. 
At $ \tau = 0$, the length of each segment is assumed to be identical, e.g., 
$S_i(0) = 1$. 
To identify the best arm, we repeat the following three processes:

\noindent {\bf (i) Selecting an arm to play}:
Select an arm with the following probability defined by 
the ratio of each segment length and total length 
$S(\tau) = \sum_{i=1}^N S_i(\tau)$ at the $\tau$-th play;
\begin{align}
 P_i(\tau) = S_i(\tau)/S(\tau) \quad \quad (i = 1, 2, ..., N).
\label{eq:def-Pi}
\end{align}
We refer to $P_i(\tau)$ as the selection probability of arm $i$ at the $\tau$-th play.

\noindent {\bf (ii) Playing the chosen arm}:
By playing the arm $i$ chosen in step (i), 
reward $x_{i}$ at the $\tau$-th play is received based 
on the reward probability distribution.

\noindent {\bf (iii) Learning and updating}: 
The length of the $i$-th segment is altered
based on the total length of the object and $x_{i}$:
\begin{align}
  S_i(\tau+1) = S_i(\tau) + b_i(x_{i},b)S(\tau),
  \label{eq:update-S_i}
\end{align}
where 
$b_{i}(x_{i},b)$ is a function of the reward $x_{i}$, 
and $b$ is a small incremental parameter.
If  $b_i(x_{i},b)<0$ and $S_i(\tau +1) <0$, $S_i(\tau +1)=0$ is set.
Although one can choose an arbitrary form of the function 
$b_{i}(x_{i}, b)$,
we use the following function in this study:
\begin{align}
b_{i}(x_{i},b) 
=\dfrac{ b \ x_{i} }{ 1 - b \ x_{i} },
\label{eq:b_i_tau} 
\end{align}
where $0< b \ll 1$ is assumed such that $bx_i < 1$. 
As the aforementioned processes (i)--(iii) are repeated, 
the length of each segment, $S_i(\tau)$, increases or decreases 
in accordance with the received reward [Fig.~\ref{fig1}(a)].
As a result, the selection probability of the best arm $i^*$ with the
largest mean reward,
$P_{i^*}(\tau) = S_{i^*}(\tau)/S(\tau)$, 
increases compared to those of the others [Fig.~\ref{fig1}(b)].

From Eqs.~(\ref{eq:def-Pi}) and (\ref{eq:update-S_i}),
the update of $P_i(\tau)$ is expressed as follows:
\begin{align}
&P_i(\tau +1) = P_i(\tau) + \Delta P_i(\tau),
\label{eq:update-P_i}
\end{align}
\begin{align}
\Delta P_i(\tau) = 
\begin{cases}
\dfrac{b_i(x_{i},b) \left(1-P_i \right)}{1+b_i(x_{i},b)}  = bx_{i}(1-P_i),
 \ \mbox{if arm $i$ is played,}
\\
-\dfrac{b_j(x_{j},b)}{1+b_j(x_{j},b)}P_i =-bx_{j}P_i, 
 \ \mbox{if arm $j$ $(\ne i)$ is played,} 
\label{eq:dP_i}
\end{cases}
\end{align}
where Eq.~(\ref{eq:b_i_tau}) was used to derive $\Delta P_i(\tau)$.

The proposed decision-making method
is analogous to
ideal gases bounded by movable partitions in a vessel.
That is, $P_i$ corresponds to the volume of the $i$-th gas 
bounded by the $(i-1)$-th and the $i$-th partitions, 
and the change in the volume of the $i$-th gas
results from the increase or decrease 
in the number of moles of the $i$-th gas corresponding to $S_i$ 
[Fig.~\ref{fig1}(b)].

The aforementioned method can also be regarded as
a modified version of the tug-of-war model \cite{Kim2010TOW,Kim2015TOW}
in the sense that the volume of each segment (probability $P_i$) 
grows and shrinks under the conservative condition of
the total volume of the body ($\sum_iP_i=1$).
Notably, the proposed decision-making method has a similarity to
the linear reward schemes, known as classical schemes in 
learning automata \cite{Baba1985NewTopicsLA,Narendra2015reviewLA},
in the case of a binary bandit problem, as well as 
replicator equations in the field of evolutionary game theory
\cite{Hofbauer2003Evolutionary}.

As described in the next section, the LV competing principle is embedded in 
the proposed decision-making method; therefore, 
an exponential decrease in the error probability 
(the probability of choosing a non-best arm) is ensured. 

\section{Results and Discussions}

\begin{figure*}[tbp]
\centering
\includegraphics[width=14cm]{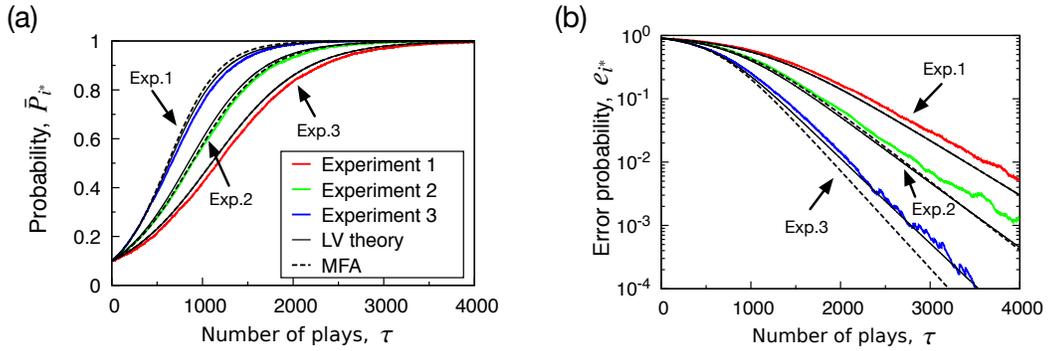}
\caption{
\label{fig2} 
(Color online) Performance evaluation for three 10-armed bandit problems.
$\tau$-dependences of (a) the mean selection probability of the best arm,
$\bar{P}_{i^*}(\tau)$,
and (b) the error probabilities in experiments 1, 2, and 3, which are 
indicated in red, green, and blue, respectively.
These results were calculated based on the update rule (i)--(iii) explained in
Sec. 2.
The reward of each arm is assumed to be provided based on a normal distribution
with mean $\mu_i$ $(i = 1, 2, \cdots ,N)$ and variance $1/4$.
In each experiment, $\mu_i$ was set as follows:
Experiment 1: $\mu_1=0.4$ and $\mu_{2\le i \le 10}=0.2$.
Experiment 2: $\mu_1=0.4$ and $\mu_{2\le i \le 10}$ were determined
from uniform random distribution in the range from $0.1$ to $0.2$.
Experiment 3: $\mu_1=0.6$, $\mu_{2\le i \le 5}=0.3$, 
and $\mu_{6\le i \le 10}=0.2$.
The black solid curves in (a) and (b) were obtained by 
numerical integration of the competitive LV equation [Eq.~(\ref{lveq})],
and the black dashed curves are analytical solutions [Eqs.~(\ref{eq:MFA-Pi})
and (\ref{eq:MFA-ei})] obtained by the mean-field approximation [see text].
}
\end{figure*}

\begin{figure}[tbp]
\centering
\includegraphics[width=7cm]{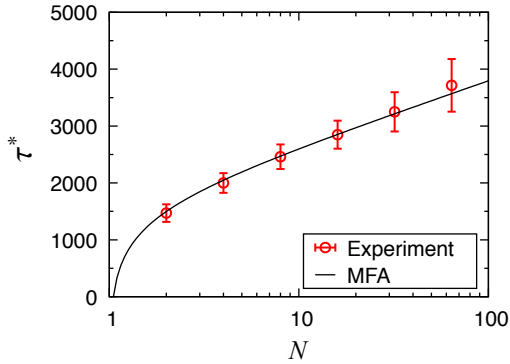}
\caption{\label{fig3} 
Average number of plays $\tau^*$ required for the error probability 
to satisfy $e_{i^*}(\tau) < \delta = 0.05$ at $N$-armed bandit problems,
where $b = 0.005$, $\mu_1=0.6$, and $\mu_{2\le i \le N}=0.2$.
The black solid curve is the theoretical estimation
obtained by the mean-field approximation (MFA) for the competitive LV equation 
[see text and Eq.~(\ref{eq:N-depend})].
}
\end{figure}

Before revealing the connection between the proposed decision-making method 
and the LV competitive dynamics,
we show the results of numerical simulations for three typical examples
to demonstrate that our method typically
has the ability to identify the best arm among multiple arms with
unknown rewards.
For the numerical demonstration, 
we set the number of arms $N=10$ and assumed that 
the reward of $i$-th arm, $x_{i}$, is provided 
based on a normal distribution, which is 
given by 
$\phi_i(x)=1/\sqrt{2\pi\sigma_i^2}\exp[-(x-\mu_i)^2/(2\sigma_i^2)]$, 
 with the mean $\mu_i$ and variance $\sigma_i^2=1/4$ 
($i=1,2,\cdots, 10$). 
The best arm $i^*$ with the largest mean reward is set to be arm
$1$, i.e., $\mu_1=\mu_{i^*}=\max_i \{\mu_i \}$.
To quantify the capability of the best arm identification, 
consecutive arm playing was conducted until the cycle $\tau = 4000$
(this makes up one ``run'').
Independently repeating this for $n_s = 100$ runs,
the mean of the selection probability of the best arm, 
$\bar{P}_{i^*}(\tau) = ({1}/{n_s}) \sum_{n=1}^{n_s} P^{(n)}_{i^*}(\tau)$, 
was evaluated, where $P^{(n)}_{i^*}(\tau)$ is the selection probability
of the best arm at the $\tau$-th play of the $n$-th run.

Figure~\ref{fig2}(a) shows the mean selection probability
of the best arm as a function of $\tau$, $\bar{P}_{i^*}(\tau)$, 
which were calculated by the update rule (i)--(iii) in Sec. 2.
As shown in this figure,
$\bar{P}_{i^*}(\tau)$ in the three typical MAB problems depicted by
red, green, and blue lines converges toward 1.
Figure~\ref{fig2}(b) shows that the error probability, 
$e_{i^*}(\tau) = 1 - \bar{P}_{i^*}(\tau)$, i.e., the probability of selecting the
non-best arm, exponentially decreases.
Hence, it is numerically shown that 
the proposed decision-making method works well in identifying the best arm
(more rigid proof of the ability of the best arm identification
will be provided in subsection \ref{sec:proof-best-arm-identification}).

The selection probability of the best arm 
is expected to depend on $N$ as well as the difference between the
mean rewards.
The average number of plays $\tau^*$ resulting 
in the decision error probability 
being sufficiently small, i.e., $e_{i^*}(\tau) < \delta = 0.05$, 
is demonstrated in Fig.~\ref{fig3} as a function of $N$.
The figure clearly shows that the average number of plays $\tau^*$
increases only with $\log N$. 
The dependence of $O(\log N)$ will be important 
in the scalable solution of MAB problems with a large number of arms, 
and it is better than the 
Upper Confidence Bound Exploration (UCB-E) \cite{Audibert2010BestArmMAB}
and an extended tug-of-war model with $O(N)$-scalability 
\cite{Naruse2018Scalable-photon}.

\subsection{Hidden Lotka--Volterra Competition Dynamics}
\label{sec:LV-derive}

In this subsection, we show that the Lotka--Volterra type of 
interspecific competition dynamics lies behind our model 
and yields
the performance of the decision-making method illustrated
in the previous subsection.

The starting point of our analysis is Eq. (\ref{eq:update-P_i})
of the selection probability of each arm.
We here consider the ensemble average of the selection probability 
of each arm at the $\tau$-th play;
$\bar{P}_i(\tau) = (1/n_{s}) \sum_{n=1}^{n_s} P_i^{(n)}(\tau)$,
where $P_i^{(n)}(\tau)$ is the selection probability at the $\tau$-th play 
on the $n$-th run.
The update of the ensemble average $\bar{P}_i(\tau)$ is described as follows:
\begin{align}
 \bar{P}_i(\tau+1)=\bar{P}_i(\tau) + \langle \Delta P_i(\tau)\rangle,
\label{eq:update-ave-P_i}
\end{align}
where $\ave{\Delta P_i(\tau)}$ is the ensemble average of $\Delta P_i(\tau)$ for
$n_s$ runs and is given by
\begin{align}
\langle \Delta P_i(\tau) \rangle 
&=
b \bar{P}_i(\tau) \left( \mu_i - \sum_{j=1}^N \mu_j \bar{P}_j(\tau) \right).
\label{eq2}
\end{align}
The detailed derivation is described in Appendix~\ref{app1}.

Next, we reconfigure ``time'' as $t \equiv \Delta t\tau = b\tau$.
When $b$ is sufficiently small, 
$\tP_i(t+\Delta t)=\tP_i(t) + (\pdev{\tP_i}{t}{})\Delta t + O\left(\Delta
t^2\right)$; thus, the average dynamics of the selection probabilities 
are described as follows:
\begin{align}
 \dev{\bar{P}_i}{t}{}
=
\bar{P_i} \left( \mu_i - \sum_{j} \mu_j \bar{P}_j \right).
\label{lveq}
\end{align}

Note that Eq. (\ref{lveq}) is equivalent to 
{\it the competitive Lotka--Volterra (LV) equation},
which describes competition among species competing 
for a common resource, as follows\cite{Baigent2016Lotka-Volterra}:
\begin{align}
 \dev{z_i}{t}{} = z_i \left( a_i - \sum_{j=1}^N c_{ij} z_j \right),
\label{lveq2}
\end{align}
where $z_i$ and $a_i$ represent the population of the species $i$ and
the growth rate of the species $i$, respectively, and
$c_{ij}$ represents intraspecific ($i=j$) and interspecific $(i\ne j)$
interactions.
By comparing Eq.~(\ref{lveq}) to Eq.~(\ref{lveq2}),
searching for the best arm in our method can be interpreted as follows:
$P_i$ attempts to ``grow'' according to $\mu_i$ 
(the mean reward of the arm $i$); however, 
the selection probabilities of the other arms also attempt to grow;
thus, they compete for survival.
The arm (species) that survives this competition will be 
considered the best arm.
The origin of this competition mechanism lies in frustration,
such as an ecosystem competing for limited resources, 
in which $P_i$ increases under the conservation conditions
of total probability; $\sum_i P_i = 1$.

\begin{figure}[tbp]
\centering
\includegraphics[width=7cm]{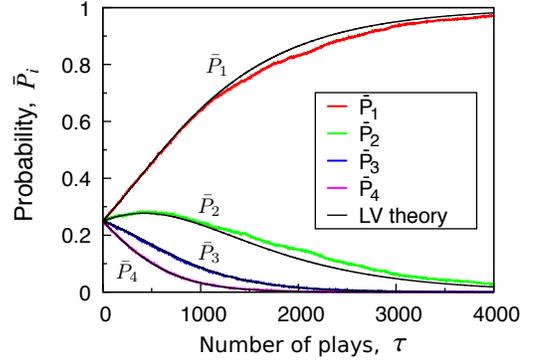}
\caption{\label{fig4} 
(Color online) $\tau$-dependence of the selection probability of arm $i$
in the 4-armed bandit problem with $\mu_i=0.6-(i-1)/10$ \ $(i=1,2,3,4)$.
The black solid curves represent the results of
the numerical integration of Eq.~(\ref{lveq}).
Via the competition, arms 2, 3, and 4 
are defeated in increasing order of the mean rewards,
and finally only arm $1$ with the largest mean reward survives.
}
\end{figure}

As an example of such competition among arms, Fig.~\ref{fig4} shows
the simulation results of the selection probability of each arm for 4-armed bandit problems,
where $\mu_1$, $\mu_2$, $\mu_3$, and $\mu_4$ are 
given by $0.6$, $0.5$, $0.4$, and $0.3$, respectively.
As a result of the competition among the arms, 
the selection probability of arm $1$ overwhelms those of the others.
The numerical results agree well with the black solid curves shown 
in Fig.~\ref{fig4}, which were
obtained by numerical integration of Eq.~(\ref{lveq})
using the Euler method with time step $\Delta t = b$.
%

\subsection{Feasibility of Best Arm Identification}
\label{sec:proof-best-arm-identification}

The competitive LV equations have been investigated 
from the context of physics and mathematics for many decades
\cite{Baigent2016Lotka-Volterra}.
Hence, accumulated knowledge and theorems regarding them can be applied to the decision-making problems examined in this study.
Among the most significant theorems known is the condition 
for only a single species to survive,
i.e., the global stability of only one solution of the competitive LV equations
\cite{Baigent2016Lotka-Volterra,Zeeman1995extinction}.
This condition ensures that the present method supported 
by the competitive LV dynamics
can eventually identify the best arm.

However, the theorem proving global stability 
is only valid in the case that all mean rewards
are different from each other ($\mu_i \ne \mu_j$).
In this subsection, 
we analyze the global stability
of a fixed point 
$\bm{P}^s=(\tP^s_1,\cdots, \tP_{i^*}^{s}, \cdots, \tP^s_N) 
= (0,\cdots,1_{i^*},\cdots, 0)$ 
in Eq. (\ref{lveq}), 
corresponding to the identification of the best arm $i^*$ with
mean probability $\tP_{i^*}=1$
and show that the fixed point is globally stable even when some mean rewards
are identical.

Here, we consider the stability by introducing a Lyapunov function
$V(\bm{P})$ for $\bm{P}=(\tP_1,\tP_2,\cdots, \tP_N)$, 
satisfying $V(\bm{P}^s)=0$ and $V(\bm{P}) >0$ for $\bm{P} \ne \bm{P}^s$, 
in a space $\Omega = \{ \bm{P} \in \mathbb{R}_{+}^N| \sum_{i=1}^N \tP_i=1 \}$.
If $\pdev{V}{t}{} <0$,
the fixed point $\bm{P}^s$ is globally stable in $\Omega$.

As the Lyapunov function, 
we chose the Kullback--Leibler (KL) divergence 
 between the probability
distribution $\bm{P}$ and
$\bm{P}^s$,
\begin{eqnarray}
V(\bm{P}) = KL(\bm{P}^s\|\bm{P})
= \sum_i^N\tP_i^s \log
\left( \dfrac{\tP^s_i}{\tP_i} \right)
= - \log\tP_{i^*}.
\end{eqnarray} 
$KL(\bm{P^s}\|\bm{P})$ represents the information gained from a prior
distribution $\bm{P}$ to a posterior distribution
$\bm{P}^s$
and satisfies $KL(\bm{P}^s\|\bm{P}) =0$ for $\bm{P} = \bm{P}^s$ 
and $KL(\bm{P}^s\|\bm{P}) >0$ for $\bm{P} \ne \bm{P}^s$. 
Similar analyses using the KL divergence as the Lyapunov function 
have been performed, for example, in evolutionary game theory \cite{Harper2011}
and switched nonlinear systems \cite{Omar2011}.
Introducing $\bar{\mu} = \sum_{i=1}^N \mu_i \tP_i$
and regarding $\tP_i^s=1$ only for $i=i^*$,
we obtain
\begin{eqnarray}
\dev{V}{t}{} = \dev{KL(\gvc{P}^s\|\gvc{P})}{t}{}
= -\dev{}{t}{} \log \tP_{i^*}
= - \left( \mu_{i^*}-\bar{\mu} \right),
\end{eqnarray}
where $\mu_{i^*}=\max_i\{\mu_i\}$ represents the maximum mean reward, 
and Eq.~(\ref{lveq}) was used.
Obviously, $\pdev{V}{t}{} = -(\mu_{i^*} - \bar{\mu}) < 0$ in $\Omega$
because of $\mu_{i^*} \ge \bar{\mu}$. 
Accordingly, we conclude that the decision-making method
described by Eq.~(\ref{lveq})
monotonically obtains the information of the best arm and the probability
$P_i^*$ to select the best arm $i^*$ converges to 1.

\subsection{Efficiency of Best Arm Identification}
\label{sec:error-prob}
This subsection provides further insight into the global behavior of $\tP_{i^*}$
by applying a mean-field approximation to Eq.~(\ref{lveq}).
Let us replace the selection probabilities, $\tP_{i \ne i^*}$, 
and the mean rewards, $\mu_{i \ne i^*}$, except for the best arm
$i^*$, with the mean values $\ave{P}$ and $\ave{\mu}$,
respectively:
$\tP_{i \ne i^*} = \ave{P} \equiv \frac{1}{N-1} \sum_{i \ne i^*} \tP_i$ and 
$\mu_{i \ne i^*} = \ave{ \mu } \equiv \frac{1}{N-1} \sum_{i \ne i^*} \mu_i$,
where $N$ is the number of arms.
In this approximation, we obtain
\begin{align}
\sum_j \mu_j \tP_j
= \mu_{i^*} \tP_{i^*} + \sum_{j \ne {i^*}} \mu_j \tP_j
= \mu_{i^*} \tP_{i^*} + (N-1)\ave{ \mu } \ave{ P }.
\label{eq:MFA-sum}
\end{align}
Furthermore, we obtain $\ave{ P } = (1-\tP_{i^*})/(N-1)$ 
by solving the relation 
$\tP_{i^*} + \sum_{j \ne {i^*}} \tP_j= \tP_{i^*} + (N-1) \ave{ P } = 1$,
where we used $\sum_j \tP_j = 1$.
Substituting the equation of $\ave{P}$ and Eq.~(\ref{eq:MFA-sum}) 
to Eq.~(\ref{lveq}), 
we obtain the mean-field LV equation as follows:
\begin{align}
\dev{\tP_{i^*}}{t}{}
=
\left( \mu_{i^*} - \ave{ \mu } \right)
\tP_{i^*} \left( 1- \tP_{i^*} \right).
\label{ave_eq}
\end{align}

Equation (\ref{ave_eq}) is easily solved, 
and $\tP_{i^*}$ and error probability,
$e_{i^*}(t) = 1 - \tP_{i^*}(t)$, are given as follows:
\begin{align}
\tP_{i^*}(t) &= \dfrac{e^{\alpha t}}{N-1 + e^{\alpha t}},
\label{eq:MFA-Pi}
\\
e_{i^*}(t) &= \dfrac{(N-1)e^{-\alpha t}}{(N-1)e^{-\alpha t}+1},
\label{eq:MFA-ei}
\end{align}
where $\alpha = \mu_{i^*}- \ave{\mu}$, and
$\tP_{i^*}(0) = 1/N$ was used as the initial selection probability.
Considering $t = b \tau $, 
it is interesting that the convergence rate does not depend on 
the number of machines, but only on $b$ and $\alpha$.

Although the above approximation is bold,
it provides reasonably good predictions.
Actually, as shown by the dashed lines in Fig.~\ref{fig2},
the time evolution given by Eqs.~(\ref{eq:MFA-Pi}) and (\ref{eq:MFA-ei})
corresponds well to the actual simulation results.

Regarding $\tau=t/b$, $0 < b \ll 1$, and $\delta \ll 1$,
one can evaluate 
the average number of plays $\tau^*$ required for the error probability 
to satisfy $e_{i^*} < \delta$ as follows:
\begin{align}
 \tau^*
&\approx 
\dfrac{1}{b(\mu_{i^*} - \ave{ \mu })} \log \dfrac{(N-1)(1-\delta)}{\delta}.
\label{eq:N-depend}
\end{align}
Thus, $\tau^*$ increases at most $ \log N $ for $N$ arms. 
The scalability of $O\left(\log N\right)$ 
well explains the numerical results shown in Fig.~\ref{fig3}.

\subsection{Adaptability to Environmental Change}

One of the most important features of reinforcement learning 
is rapid and robust adaptation to non-stationary environments, 
in which the mean rewards change over time.
The correspondence between our decision-making method and 
the competitive LV equations suggests an insight into 
how quickly the system adapts 
to environmental change.
In Appendix~\ref{sec:adaptability},
the adaptability of our decision-making method to an environmental change
is discussed in terms of natural biodiversity.
%

\section{Summary}

In this study, we developed a decision-making principle for solving MAB
problems, in which the best choice or arm identification 
is theoretically guaranteed by the LV competitive mechanism.
Furthermore, by applying the mean-field approximation 
to the competitive LV equations, we showed 
that the error probability exponentially decreases and that 
the time required for the best arm identification 
depends on only a logarithm of the number of arms, which is an important
attribute in realizing decision-making scalability. 

We remark that the aforementioned consideration is valid when $b\ll 1$. 
The estimation of the appropriate $b$-value for
the correspondence between the update rule (i)--(iii) and the
LV equation [Eq.~(\ref{lveq})] may be a difficult problem
because the appropriate $b$-value may 
depend on the reward probability distributions.
The detailed investigation will be interesting future work.

The present study of our decision-making method demonstrates the possibility 
of using competitive systems for reinforcement learning.
Methods harnessing nature may incorporate the
superior performance of natural processes in adaptability and robustness,
and provide a means to map our knowledge of nature 
into reinforcement learning techniques. 

\section*{Acknowledgments}
This work was supported in part by the Japan Science
and　Technology  Agency  CREST  Grant  Number  JPMJCR17N2,
PRESTO Grant Number JPMJPR19M4, 
Japan  Society  for  the  Promotion of Science Grants-in-Aid for
Scientific Research Grant No. JP17H01277, JP19H00868,
and Murata Science Foundation.

\appendix 

\section{Derivation of $\ave{\Delta P_i(\tau)}$}
\label{app1}

To evaluate the average amount of change in $P_i(\tau)$
that varies according to the results of a probabilistic trial,
we consider the ensemble average of the selection probability for $n_s$
runs, as described in the main text.
Here, let $P_i^{(n)}(\tau)$ be the selection probability of arm $i$ at
the $\tau$-th play on the $n$-th run. 
The ensemble average at the $\tau$-th play is defined as
$
\tP_i(\tau) = (1/n_{s}) \sum_{n=1}^{n_s} P_i^{(n)}(\tau)
$ 
in the limit of $n_s \rightarrow \infty$.

Recalling the update rule of our method described in Eq.~(\ref{eq:dP_i}),
we describe the ensemble average of $\Delta P_i^{(n)}(\tau)$ 
at the $\tau$-th play for $n_s$ runs as follows:
\begin{align}
\langle \Delta P_i(\tau) \rangle &= 
\dfrac{1}{n_{s}}\sum_{n=1}^{n_{s}}\Delta P_i^{(n)}(\tau) \nonumber \\
&=
\dfrac{1}{n_{s}}
\left\{ \sum_{n=1}^{n_{i}}
  b x^{(n)}_{i,\tau} \left( 1-P_i^{(n)}(\tau) \right)
  - \sum_{j \ne i} \sum_{n=1}^{n_{j}}
  b x^{(n)}_{j,\tau} P_i^{(n)}(\tau)
\right\} 
\nonumber \\ &=
b\dfrac{n_i}{n_{s}} \dfrac{1}{n_{i}}
\sum_{n=1}^{n_{i}} x^{(n)}_{i,\tau} \left( 1-P_i^{(n)}(\tau) \right)
-
b\sum_{j\ne i} \dfrac{n_j}{n_{s}} \dfrac{1}{n_{j}}
\sum_{n=1}^{n_{j}} x^{(n)}_{j,\tau} P_i^{(n)}(\tau),
\label{eq:averae-P_i}
 \end{align}
where $n_i$ is the number of runs that arm $i$ is selected at the
$\tau$-th play, 
and $x_{i,\tau}^{(n)}$ is a reward obtained from arm $i$ at the $\tau$-th play
on the $n$-th run.
Note that $x_{i,\tau}^{(n)}$ is stochastically determined
according to the probability distribution with the mean reward $\mu_i$ of the arm $i$ 
and is independent of $P_i^{(n)}(\tau)$.

Thus, the following relationship holds when $n_i$ is sufficiently large:
\begin{align}
 \dfrac{1}{n_{i}} \sum_{n=1}^{n_{i}} x^{(n)}_{i,\tau}P_i^{(n)}(\tau)
\approx
\mu_i \bar{P}_i(\tau).
\label{eq:approx_mu_P_i}
\end{align} 
Because the ratio $n_i/n_s$ means that the arm $i$ is selected $n_i$ times 
out of $n_s$ runs at the $\tau$-th play, the ratio can be interpreted as a mean value 
of the selection probability:
\begin{align}
  \dfrac{n_i}{n_s} = \bar{P}_i(\tau).
\label{eq:average-P_i}
\end{align}
Substituting Eqs.~(\ref{eq:approx_mu_P_i}) and (\ref{eq:average-P_i})
to Eq.~(\ref{eq:averae-P_i}),
we can obtain
\begin{align}
\langle \Delta P_i(\tau) \rangle 
&\approx 
b\bar{P}_i(\tau)
\left\{  \mu_i(1-\bar{P}_i(\tau))  \right\}
- b\sum_{j\ne i} \mu_j\bar{P}_j(\tau)\tP_i(\tau)
\nonumber \\&=
b\bar{P}_i(\tau) \left( \mu_i - \sum_{j=1}^N\mu_j \bar{P}_j(\tau) \right)
\end{align}
in the limit of $n_s \rightarrow \infty$.

%
\section{Simulations of Adaptability}
%
\label{sec:adaptability}

As a demonstration of the adaptability of 
the presented decision-making method to non-stationary environments,  
we numerically performed the simulations of the MAB problem ($N=4$, $b=0.01$),
in which the mean rewards are 
cyclically changed at a constant interval of $T=15000$ steps as follows:
$(\mu_1, \mu_2, \mu_3, \mu_4)=$ $(0.6,0.5,0.4, 0.3)$ $\to$
$(0.3,0.6,0.5,0.4)$ $\to$ $(0.4,0.3,0.6,0.5)$ $\to$ $(0.5,0.4,0.3,0.6)$
$\to$ $(0.6,0.5,0.4, 0.3)$.
To achieve adaptable decision-making in the simulations, 
we introduced a lower bound of the selection probability of each arm,
$P_{\text{min}}$, such that $P_i \ge P_{ \text {min}}$
no matter how much the selection probability decreases 
as a result of the search.
In this study, $P_{\text{min}}$ is used as 
a control parameter regarding the decision-making adaptability
and optimality. 

\begin{figure}[tbp]
 \centering

\includegraphics[width=7cm]{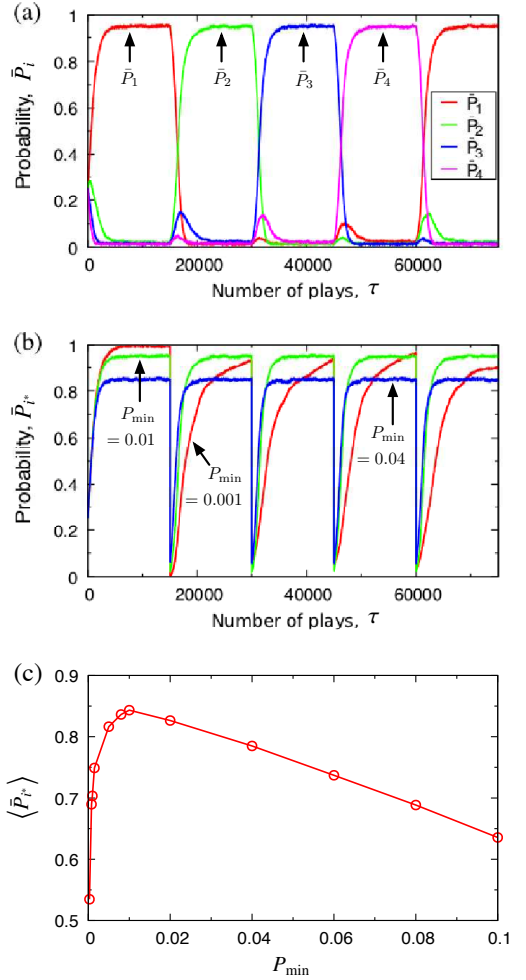}
\caption{
(Color online) Simulation results of the 4-armed bandit problem
in a non-stationary environment where the mean rewards are 
cyclically changed at a constant interval of $T=15000$ steps.
(a) Selection probability of arm $i$, where $P_{\text{min}} = 0.01$ was
 used.
The best arm is swapped in response 
to periodic changes in the mean reward.
(b) Selection probability of the best arm, $\tP_{i^*}$,
where the red, green, and blue curves
represent the results for $P_{\text{min}} = 0.001$, $0.01$, and $0.04$,
respectively.
(c) Relation between the mean of $\ave{\tP_{i^*}}$ and $P_{\text{min}}$.
}
\label{fig:adaptability}
\end{figure}

The time evolution of the average selection probability of each arm, $\tP_i$,
when $P_{\text{min}} = 0.01$ is shown in Fig.~\ref{fig:adaptability}(a), where
the best arm is switched every $15000$ steps.
Immediately after the environmental changes,
competition has arisen, and then, 
the system eventually finds the best arm during each term.

Fig.~\ref{fig:adaptability}(b) shows the time evolution of the selection
probability of the best arm, $\tP_{i^*}$, with different values of
$P_{\text{min}}$.
The maximum value of $ \tP_{i^*}$ can be approximately given as
$1- (N-1) P_{\text{min}}$, resulting in low optimality for a too large 
$P_{\text{min}}$, 
whereas a too small $P_{\text{min}}$ makes the adaptive best arm
identification difficult. 
To investigate the balance between the optimality and adaptability,
we calculated the mean value of $\tP_{i^*}$ over time from $2T$ to $5T$:
\begin{align}
 \ave{{\tP}_{i^*}} = \frac{1}{3T} \int^{5T}_{2T} \tP_{i^*}(t) \ \intd{t}.
\end{align}
As can be seen in Fig.~\ref{fig:adaptability}(c), 
a smaller value of $P_{ \text{min}}$ produces a larger $\ave{{\tP}_{i^*}}$,
but in the range where $P_{ \text{min}} < 0.01$,
$\ave{{\tP}_{i^*}}$ rapidly decreases.
Thus, under this setting, $\ave{{\tP}_{i^*}}$ has a peak value 
of approximately $P_{\text{min}} = 0.01$.

The simulation results in this appendix can be interpreted as follows:
the adaptability is maximized
by preventing extinction of all species, even though some of them are not optimal
in a certain environment, from the perspective of the
``winner-take-all'' 
competition mechanism in ecosystems. 
This interpretation is reminiscent of the ecosystem stability exerted by
natural biodiversity (species richness).
Though the actual effect of biodiversity on the stability of ecosystems
is more complicated\cite{Pennekamp2018Biodiversity},
this type of analogy might provide a new insight into the field of 
reinforcement learning.

\bibliographystyle{jpsj}

\end{document}